\documentclass[aps,floats,epsf,twocolumn]{revtex4}
 \usepackage{graphics}

\begin{document}

\title{Zero-energy states and fragmentation of spin in
the easy-plane antiferromagnet on a honeycomb lattice}

\author{Igor F. Herbut}

\affiliation{Department of Physics, Simon Fraser University,
 Burnaby, British Columbia, Canada V5A 1S6}

\begin{abstract}

The core of the vortex in the N\'{e}el order parameter for an easy-plane
antiferromagnet on  honeycomb lattice is demonstrated to bind two
zero-energy states. Remarkably, a single electron occupying this
mid-gap band has its spin fragmented between the two sublattices:
Whereas it yields a vanishing total magnetization it shows a finite
N\'{e}el order, orthogonal to the one of the assumed background. The
requisite easy-plane anisotropy may be introduced by a magnetic
field parallel to the graphene's layer, for example. The results are
relevant for spin-1/2 fermions on graphene's or optical honeycomb
lattice, in the strongly interacting regime.

\end{abstract}
\maketitle

\vspace{10pt}

 The fact that the low-energy quasiparticles in graphene behave as massless Dirac fermions has placed this new
incarnation of carbon-based structures at the center stage of
condensed-matter physics \cite{castroneto}. Although in its natural
state graphene appears to be a gapless semimetal, it is tempting to
also  consider it in its possible strongly correlated phases, where
the quasiparticles would become gapped and the ground state
insulating. Such a semimetal-insulator transition in graphene would
in fact represent a condensed-matter analog of the Higgs mechanism,
with the Higgs field as a quasiparticle composite
\cite{khveshchenko}, \cite{herbut1}. Although the ultimate nature of
the insulating phase that would result from the strong Coulomb
interaction may depend on the details at the scale of the lattice
constant, the simplified Hubbard model on a honeycomb lattice is
most likely to have the standard N\'{e}el-ordered ground state at a
sufficiently large $U/t$. This is certainly what is expected at
$U/t=\infty$, and is also supported by numerical \cite{sorella} and
analytical \cite{herbut1} studies, already for $U/t$ above some
critical value.

As long as the quantum phase transition between the semimetal and
the antiferromagnetic insulator is not strongly discontinuous and  not
too deep in the insulating phase, it is still sensible to consider
fermions as itinerant. Weak antiferromagnetic ordering is then felt
by the formerly gapless quasiparticles as an opening of a
``relativistic" mass gap, with the ``mass" being proportional to the
N\'{e}el order parameter. The unusual universality class of this
transition is completely determined by the presence of gapless Dirac
fermions on the semimetallic side \cite{herbut1}. It seems natural
to ask if there could be other interesting consequences of this fact
in the insulating phase itself. In particular, in light of the known
properties of the spectrum of Dirac's Hamiltonian in topologically
non-trivial backgrounds \cite{jackiw}, \cite{hou} one wonders if
topological defects in the N\'{e}el order here may localize sub-gap
states at zero energy. Such states have been known to display a
number of unusual features, such as charge fractionalization
\cite{jackiw}, \cite{hou} and non-abelian statistics \cite{read} and
are being discussed intensively in the context of topological
quantum computing \cite{oshikawa}.

I show here that this is indeed the case at least when there is an easy-plane
anisotropy for the N\'{e}el order, when the core of the vortex in
the order parameter turns out to contain a pair of such mutually
orthogonal zero-energy states. The electric charge of the vortex
in this case is integer; however, the core states have the positional and
spin degrees of freedom entangled in an interesting way. At
half filling, this two-state mid-gap band is half full. A single
electron occupying the zero-energy band then has the direction of
its spin alternating between the two triangular sublattices of the
honeycomb lattice. Its spin after averaging over the unit cell
therefore vanishes. Its N\'{e}el order parameter, however, is
finite, and points in the direction orthogonal to the easy-plane. An
electron occupying the zero-energy state would thus represent a
local {\it single-particle antiferromagnet}. Placing the system in a
weak magnetic field parallel to the graphene layer is further
demonstrated to introduce such an easy-plane anisotropy for the
N\'{e}el ordering. This may help create the experimental conditions
for an observation of this effect in graphene's or optical honeycomb
lattice.

Let us begin by defining the standard Hubbard model as $H_t + H_U$,
with
\begin{equation}
H_t= -t \sum_{\vec{A}, i, \sigma=\uparrow, \downarrow} u^\dagger
_\sigma (\vec{A}) v_\sigma (\vec{A}+\vec{b}_i) + H. c.,
\end{equation}
\begin{equation}
H_U = U \sum_{\vec{X}} n_\uparrow (\vec{X}) n_{\downarrow}
(\vec{X}).
\end{equation}
The sites $\vec{A}$ denote one triangular sublattice of the
hexagonal lattice, generated by linear combinations of the basis
vectors $\vec{a}_1= (\sqrt{3}, -1)(a/2)$, $\vec{a}_2 = (0,a)$. The
second sublattice is then at $\vec{B}=\vec{A}+\vec{b}$, with the
vector $\vec{b}$ being either $\vec{b}_1= (1/\sqrt{3},1) (a/2)$,
$\vec{b}_2= (1/\sqrt{3},-1) (a/2)$, or $\vec{b}_3= (-a/\sqrt{3},0)$.
$a$ is the lattice spacing, and $U>0$.

The spectrum of $H_t$ for each projection of spin becomes linear in
the vicinity of the two non-equivalent Fermi points, which we choose
to be  at $\pm \vec{K}$, with $\vec{K} = (1,1/\sqrt{3}) (2\pi/a
\sqrt{3})$ \cite{semenoff}. In vicinity of these points the
linearized Hamiltonian assumes a ``relativistically invariant" form
$H_0= \gamma_0 \gamma_i \partial_i$ with $\gamma_0 = I_2 \otimes \sigma_3$, $\gamma_1=
\sigma_3 \otimes \sigma_2$, $\gamma_2= I_2 \otimes \sigma_1$, where
$\{ I_2, \vec{\sigma} \}$ is the standard Pauli basis. The global
``chiral" SU(2) symmetry for each spin projection is generated by
$\{ \gamma_{35}, \gamma_3, \gamma_5 \}$, where $\gamma_3= \sigma_1
\otimes \sigma_2$, $\gamma_5= \sigma_2 \otimes \sigma_2$, and
$\gamma_{35}= i \gamma_3 \gamma_5= \sigma_3 \otimes I_2$
\cite{herbut3}. In the continuum limit, the four-component
wave-function for each spin projection is given by $\Psi_\sigma
^\dagger (\vec{x}) = (u^\dagger _{1\sigma} (\vec{x}), v^\dagger
_{1\sigma} (\vec{x}),
 u^\dagger _{2\sigma} (\vec{x}), v^\dagger _{2\sigma}
(\vec{x}) )$, where
\begin{equation}
u_{i\sigma}(\vec{x})= \int^\Lambda \frac{d\vec{q}}{(2\pi a )^2} e^{
- i \vec{q}\cdot \vec{x}} u_{1\sigma} ((-)^{i+1} \vec{K}+\vec{q}),
\end{equation}
and analogously for $v_{i \sigma}(\vec{x})$. The ``valley" index
$i=1,2$  labels the two Fermi points. The reference frame has been
rotated so that $q_x = \vec{q}\cdot \vec{K}/K$ and $q_y =
(\vec{K}\times \vec{q})\times \vec{K}/K^2 $, and $\hbar=k_B=v_F =1$,
where $v_F =ta\sqrt{3}/2$ is the Fermi velocity.

The interaction term $H_U$ can also be written as
 \begin{eqnarray}
H_U= \frac{U}{16} \sum_{\vec{A}} [ (n(\vec{A} ) + n(\vec{A}+\vec{b})
)^2 + (n(\vec{A} ) - n(\vec{A}+\vec{b}) )^2 \\ \nonumber
-(\vec{f}(\vec{A} ) + \vec{f}(\vec{A}+\vec{b}) )^2 - (
\vec{f}(\vec{A} ) - \vec{f} (\vec{A}+\vec{b}) )^2 ],
\end{eqnarray}
where  $n(\vec{A})  = u^\dagger_\sigma (\vec{A})  u _{\sigma}
(\vec{A}) $ and $\vec{f} (\vec{A})  = u^\dagger_\sigma (\vec{A})
\vec{\sigma}_{\sigma \sigma'}  u _{\sigma'} (\vec{A}) $, are the
particle number and the magnetization operators at the site
$\vec{A}$. Variables at the second sublattice are analogously
defined in terms of $v_\sigma (\vec{B}) $. This is the rotationally
invariant version of the  decomposition previously introduced in
\cite{herbut1}.

Assume next that the ground state has a uniform average density,
zero average magnetization, and finite N\'{e}el order: $ \langle n
(\vec{X})\rangle = n$, $\langle \vec{f}( \vec{A} ) + \vec{f}
(\vec{A}+\vec{b})\rangle=0$, $ \vec{N} = \langle \vec{f}( \vec{A} )
- \vec{f} (\vec{A}+\vec{b})\rangle \neq 0$.
 For the Hubbard model we
believe this to be the leading instability of the semi-metallic
ground state above critical $U/t$ \cite{herbut1}. By the usual
Curie-Weiss decoupling the last term in Eq. (4) yields then the
mean-field Hamiltonian for the Dirac fermions in a fixed N\'{e}el
background $\vec{N}(\vec{x})$:
\begin{equation}
H_{N}= I_2 \otimes H_0 - (\vec{N} (\vec{x}) \cdot \vec{\sigma})
\otimes \gamma_0,
\end{equation}
with the first operator in the direct product acting on the spin,
and the second on the Dirac indices. For example, if
$\vec{N}(\vec{x}) = (0,0,N)$ and constant, in the ground state of
$H_N$ the average number of spin-up (spin-down) electrons is larger
on the sublattice B (A). The ground state of $H_{N}$ exhibits
therefore a uniform N\'{e}el ordering along the third spin axis in
this case. Since all three $\gamma$-matrices appearing in $H_0$ are
diagonal in valley indices, one can rewrite $H_{N}= H_1 \oplus H_2$,
with
  \begin{equation}
  H_{1(2)} = \pm I_2 \otimes \sigma_1 (-i \partial_1) + I_2 \otimes \sigma_2
  (i \partial_2) - (\vec{N}(\vec{x}) \cdot \vec{\sigma})\otimes
  \sigma_3
  \end{equation}
as the intra-valley Hamiltonians near the Fermi points at $\vec{K}$
($H_1$), and $-\vec{K}$ ($H_2$). Both $H_1$ and $H_2$ on the other
hand may be recognized as unitary transformations of the generic
Hamiltonian $H$: $H= H_0 + \vec{m} (\vec{x}) \cdot \vec{M}$, where
$\vec{M} = ( i \gamma_0 \gamma_3, i\gamma_0 \gamma_5,\gamma_0)$.
Specifically, $H_1 = U_1 ^\dagger H  U_1$, with
\begin{equation}
U_1 = I_2 \oplus i \sigma_2
\end{equation}
and $ \vec{m}(\vec{x})=( N_1(\vec{x}), N_2(\vec{x}),
-N_3(\vec{x}))$, and $H_2 = U_2 ^\dagger H U_2$, with
\begin{equation}
U_2 =i \sigma_2 \oplus I_2
\end{equation}
and $ \vec{m}(\vec{x})=( N_1(\vec{x}), N_2(\vec{x}), N_3(\vec{x})
)$. $H$ represents the most general single-particle Hamiltonian in
two dimensions with the relativistic spectrum $ E^2 = k^2 + m^2$ and
with the chiral symmetry of $H_0$ broken \cite{herbut2}. It may also
be understood as the mean-field Hamiltonian for spinless particles
on a honeycomb lattice in the background of the order parameter
$\vec{m}$ $= \langle \Psi^\dagger \vec{M} \Psi \rangle$. For
instance, $\vec{m}=(0,0,m)$ is the familiar state with a density
imbalance between the A and B sublattices \cite{semenoff},
\cite{haldane}, \cite{herbut1}, whereas $\vec{m}=(m_1,m_2,0) $
represents a state with broken translational symmetry, with the
``Kekule" pattern of hopping integrals of different magnitudes
\cite{hou}.

Finding the spectrum of $H_{N}$ is therefore a special case of the
general problem of diagonalizing $H$ for a given configuration of
the mass matrix $\vec{m}(\vec{x})$. Here I
   focus on the issue of zero-energy states for the vortex configuration.
 Assume that one of the components of $\vec{m} (\vec{x})$ vanishes
 everywhere, say $m_3 (\vec{x})=0$. As a result, $\{\gamma_0, H \}=0$. Since $\gamma_0 ^2
 =1$, $\gamma_0=P_+ - P_-$ where $P_\pm$ are the projectors onto
two orthogonal eigenspaces corresponding to  $\pm 1$ eigenvalues.
This, on the other hand, means that when $m_3 (\vec{x})=0$, $H = P_+
H P_- + P_- H P_+ $, so that $H$ is block-off-diagonal in the
eigenbasis of $\gamma_0$ in which
 $ P_+ = I_2 \oplus 0 $  \cite{thaller}.  For $\gamma_0$  as defined right below Eq. (1) this change of basis
  amounts to a simple exchange of the second and the third components of $\Psi (\vec{x})$.  Next, assume that
  $\Delta = m_1 (\vec{x})+ i m_2 (\vec{x})=|\Delta(r)| e^{i p
  \phi}$, $p=\pm 1$, is the vortex configuration, with $(r,\phi)$ as polar coordinates
and, and $|\Delta(r\rightarrow \infty )|\rightarrow m$. This problem
was considered by Jackiw and Rossi \cite{jackiw2} and recently by
Hou et al \cite{hou}. Defining $\partial_z =\partial_x+ i
\partial_y$, the zero-energy state $\Psi_0 (\vec{x})$ satisfies either
 \begin{equation}
 i\partial_z v (\vec{x}) + \Delta^* v^* (\vec{x})  =0,
 \end{equation}
 with $v_1=v_2 ^* =v(\vec{x})$ and $u_1=u_2=0$, or
\begin{equation}
 i\partial_{\bar{z}} u (\vec{x}) + \Delta^* u^* (\vec{x})=0
 \end{equation}
with $u_1 = u_2 ^* =u(\vec{x})$, and $v_1=v_2=0$.  For $p=1$ one
finds:
\begin{equation}
v(\vec{x})= \frac{C}{r}  e^{ -i \phi} f(r),
\end{equation}
 with $u=0$ and $-\ln f(r)= \int_0 ^r |\Delta(t)| dt + i (\pi/4) $ , and
\begin{equation}
u(\vec{x})= C f(r)
\end{equation}
with $v=0$. $C$ is the normalization factor. Only the second
solution is normalizable, however, and thus represents the unique
zero-energy eigenstate of $H$. The situation is reversed for an
antivortex, when $p=-1$. For general $p$ one can show that there are
$|p|$ linearly independent zero-energy bound states.

  An important general property of the zero-energy states should
  be noted. Consider the following sum:
  \begin{equation}
  q(\vec{x})= \sum_{E\in R}  \Psi_E ^\dagger  (\vec{x}) Q
  \Psi_E  (\vec{x}),
  \end{equation}
where $Q$ is a traceless Hermitean matrix, and $\{ \Psi_E (\vec{x})
\}$ the eigenstates of the generic Hamiltonian $H$. If the summation
is performed over the whole spectrum, $q(\vec{x})\equiv 0$. If $R$
includes only the occupied states, on the other hand, the sum
represents the ground-state average of a physical observable,
$\langle q(\vec{x}) \rangle$. By subtracting a half of the vanishing
sum over the whole spectrum one can rewrite this same average as
\cite{gordon}
\begin{equation}
\langle q(\vec{x}) \rangle =  \frac{1}{2} ( \sum_{occup}-
\sum_{unoccup}) \Psi_E ^\dagger (\vec{x}) Q \Psi_E (\vec{x}).
\end{equation}
The last form makes it evident that if there exists a unitary matrix $T$
that anticommutes with $H$, for any $Q$ that commutes 
with $T$ the only states that can contribute to $\langle q(\vec{x})
\rangle $ are the states with zero energy. For the vortex
Hamiltonian considered above, $T=\gamma_0$. Choosing $Q=\gamma_0$
yields then $\langle q(\vec{x}) \rangle= \langle m_3 (\vec{x})
\rangle $, whereas for $Q=i\gamma_0\gamma_3$, $\langle q(\vec{x})
\rangle= \langle m_1 (\vec{x}) \rangle $ for example. The important
observation is that the  contribution to the hard-axis component of
the order parameter comes {\it exclusively} from the zero-energy
state.

We now translate these results to the vortex in the N\'{e}el order.
Take the N\'{e}el vector to be in the plane 1-2: $\vec{N} = (N_1
(\vec{x}), N_2 (\vec{x}), 0)$, with $\Delta (\vec{x})= N_1
(\vec{x})+i N_2 (\vec{x})= |\Delta (r) | e^{i\phi}$. The zero-energy
eigenvalue of $H_1$ is then
\begin{equation}
\Psi_{1,0}= \left( \begin{array}{c} u_{1\uparrow} \\
v_{1\uparrow}
\\ u_{1\downarrow} \\ v_{1\downarrow} \end{array} \right)
= U_1^\dagger \Psi_0
= f(r)   \left( \begin{array}{c} 1\\ 0\\
0\\i \end{array} \right ),
\end{equation}
and of $H_2$,
\begin{equation}
\Psi_{2,0}= \left( \begin{array}{c} u_{2\uparrow} \\
v_{2\uparrow}
\\ u_{2\downarrow} \\ v_{2\downarrow} \end{array} \right)
= U_2^\dagger  \Psi_0
= f(r) \left( \begin{array}{c} 0\\ 1\\
i\\0 \end{array} \right ),
\end{equation}
with $f(r)$ assumed to be normalized. The two zero-energy states are
degenerate, orthogonal, and exponentially localized at the center of
the vortex. Consider a linear combination
   $\Psi^\dagger = a^* (\Psi_{1,0} ^\dagger ,0)+ b^* (0,\Psi_{2,0}^\dagger) $,
$ |a|^2 + |b|^2 = 1 $. In this state,
   \begin{equation}
   u_\uparrow (\vec{x})= a e^{i\vec{K}\cdot\vec{x}} f(r) ,
   \end{equation}
    \begin{equation}
   u_\downarrow (\vec{x}) =
    i b  e^{-i\vec{K}\cdot\vec{x}} f(r) ,
   \end{equation}
   and
\begin{equation}
   v_\uparrow (\vec{x})= b  e^{-i\vec{K}\cdot\vec{x} } f(r),
   \end{equation}
    \begin{equation}
   v_\downarrow (\vec{x}) = i a  e^{i\vec{K}\cdot\vec{x} } f(r).
   \end{equation}
where $w_\sigma = w_{1\sigma} e^{i\vec{K}\cdot\vec{x}} + w_{2\sigma}
   e^{-i\vec{K}\cdot\vec{x}}$, $w=u,v$. Introducing $u^\dagger = (u_\uparrow ^*, u_\downarrow ^*)$
 and $v^\dagger = (v_\uparrow ^*, v_\downarrow ^*)$, the average particle densities on the sublattices A and B
 in the state $\Psi$ are equal to
 \begin{equation}
 u^\dagger u = v^\dagger v = | f(r)|^2,
 \end{equation}
independently of the state $\Psi$. The average of the third
component of the spin, on the other hand, is
\begin{equation}
 u^\dagger \sigma_3  u = - v^\dagger \sigma_3 v=
 (|a|^2 - |b|^2) | f (r)|^2,
 \end{equation}
and thus {\it alternating} between the two sublattices. For the
easy-plane components of the average spin we can compactly write,
\begin{equation}
-i u^\dagger \sigma_+ u = i v^\dagger \sigma_-  v = 2 a^* b e^{-2
i\vec{K}\cdot\vec{x}} | f(r)|^2,
\end{equation}
which is evidently zero for  both $\Psi=\Psi_{1,0}$ and
$\Psi=\Psi_{2,0}$ \cite{comment}.

Exactly at half-filling the two-state band at zero-energy is
occupied by a single electron, which is in some arbitrary state
$\Psi$. Whereas in any such state the electron has an equal
probability to be found on either sublattice, the projection of its
spin along {\it any} axis when integrated over the whole system is
zero \cite{comment1}. However, unless $|a|=|b|$, the electron spin
will manifest itself in the {\it staggered} magnetization in the
direction orthogonal to the plane of the N\'{e}el background. Taking into
account the zero-states' local contribution to the order parameter
the vortex would thus be turned into a  half-skyrmion. It is
a remarkable property of the zero-energy states that the spin of a
single electron is distributed in such an alternating pattern. It
agrees, however, with the observation on the general nature of the
zero-energy states made earlier.

As shown below, one way of introducing such an easy plane for the
N\'{e}el order parameter is to turn on a uniform magnetic field and
couple it to particle's spin. It is of interest therefore to
understand how that would perturb the zero-energy states. In the
zero-energy subspace spanned by $(\Psi_{1,0} ^\dagger,0) $ and
$(0,\Psi_{2,0} ^\dagger)$ the Zeeman term is represented by the
perturbation
\begin{equation}
H_Z = \lambda (\sigma_3 \otimes I_2) \oplus  (\sigma_3 \otimes I_2),
\end{equation}
where $\lambda = -\mu_B B$. It is obvious  that $\langle \Psi_{i,0}
| H_3 | \Psi_{j,0}\rangle =0 $, and there is no shift in energies to
the first order in $B$. Physical reason is precisely the states'
vanishing magnetization, which decouples them from the magnetic
field to the first order.

The energies do shift, however,  if the N\'{e}el vector is tilted
out of the easy plane. Assume a configuration $\vec{N} = (N_1
(\vec{x}), N_2 (\vec{x}), N_3)$, with $N_1 (\vec{x}) + i N_2
(\vec{x})= |\Delta(r)| e^{i\phi}$ as before, but with a uniform $N_3
\ll |\Delta(\infty)| $. Within the two-dimensional zero-energy
subspace this perturbation is represented by
  \begin{equation}
  H_3 = -N_3 (\sigma_3 \otimes \sigma_3 ) \oplus (\sigma_3 \otimes
  \sigma_3).
\end{equation}
This gives $\langle \Psi_{i,0} | H_3 | \Psi_{j,0}\rangle =\pm N_3
\delta_{ij} $. A finite component $N_3$ at half filling would
therefore force one of the states in Eqs. (15)-(16) to be occupied
and the other one to be empty.

Let us show how the Zeeman coupling to the magnetic field introduces
easy-plane anisotropy for the N\'{e}el order in the present case. To
this purpose add the term $\lambda (\sigma_3 \otimes I)$ to $H_{N}$
in Eq. (5), and assume a uniform $\vec{N}$. The third spin axis, of
course, may be in an arbitrary direction in the real space.
Decomposing the new Hamiltonian again as right above  Eq. (6), one
now finds $H_1$ and $H_2$ as unitary transforms of $H + \lambda
\gamma_{35}$, with the eigenstates \cite{so3}:
\begin{equation}
E(k) = \pm [ (\sqrt{ N_3 ^2 + k^2 }\pm |\lambda| )^2  + (N_1 ^2 +
N_2 ^2) ] ^{1/2}.
\end{equation}
In the ground state at half filling all the negative energy states
are occupied and the positive energy states empty. It is easy to
show then that among all possible orientations of $\vec{N}$ the
lowest energy belongs to the one with $N_3=0$. An obvious way to
enforce  such a purely Zeeman coupling would be to orient the
magnetic field parallel to the plane of the honeycomb lattice. The
N\'{e}el order parameter would then be confined to the plane
orthogonal to the honeycomb lattice.

Hou et al. \cite{hou} have recently found zero-energy states on
graphene's honeycomb lattice in the spectrum of spinless electrons
in the vortex in their two-component ``Kekule" order parameter, as
discussed below Eq. (8). In that case there is a single zero-energy
state, a remarkable consequence of which is that the charge of a
vortex that binds it is fractionalized. In contrast, since we have
two zero-energy states, the charge of each is simply unity, but it
is the spin properties of the states that are nontrivial. The
crucial feature of our example, however, is that the relevant $U(1)$
symmetry of the order parameter is the exact rotational symmetry.
This makes one cautiously optimistic about the survival of the
mid-gap states once the lattice effects are fully restored. In this
context it would be particularly interesting to diagonalize  the
lattice version of $H_N$ numerically. A related recent calculation \cite{seradjeh}
shows zero-energy states to be surprisingly resilient to the effects of the
 discrete lattice.

  In conclusion, the
  mean-field Hamiltonian for Dirac quasiparticles in the background
  of a vortex in weak N\'{e}el order on graphene's honeycomb lattice
  has two orthogonal core states at zero energy, with positional and
  spin degrees of freedom maximally entangled.
  An electron in one of these states at half filling
  is a single-particle antiferromagnet with a finite, localized
  contribution to the staggered magnetization in the direction
  orthogonal to the easy plane, and zero average magnetization.
  The required easy-plane anisotropy was shown to
  be introduced by placing the graphene layer in a weak parallel
  magnetic field.

  This work was supported by NSERC of Canada. The author is also
grateful to KITP at UC Santa Barbara (NSF grant PHY99-07949) for
hospitality during its graphene workshop, and to C. Chamon and M.
Franz for critical reading of the manuscript and useful discussions.


\begin{thebibliography}{99}
\bibitem{comment} Furthermore, since the continuum limit is meaningful only if
$|\Delta(r)|$ is a slowly varying function on the scale of the
lattice spacing, the easy-plane components in Eq. (23) vanish even
if $a^* b \neq 0$ due to the rapidly oscillating prefactor. As the
single length scale in $H$ is $v_F/m$, this requires the mass-gap to
be small compared to the bandwidth, and the system thus to be close
to a (nearly) continuous semimetal-insulator phase transition.
\end{thebibliography}
\end{document}